# Ultra-Fast Wireless Power Hacking

Hui Wang, Hans D. Schotten, and Stefan M. Goetz

**Abstract**—The rapid growth of electric vehicles (EVs) has driven the development of roadway wireless charging technology, effectively extending EV driving range. However, wireless charging introduces significant cybersecurity challenges. Any receiver within the magnetic field can potentially extract energy, and previous research demonstrated that a hacker could detect the operating frequency and steal substantial power. However, our approach required time to track new frequencies or precise adjustments of inductance and capacitance, which would be less effective against potential rapid transmitter frequency changes or capacitance drift. As a solution, we enhanced the interceptor and enabled it to intrude as well as steal energy within just three cycles of the high-frequency signal. Moreover, it can work without any circuit parameters or look-up tables. The key innovation is synchronizing the receiver current with the phase of the magnetic sensor voltage. Through MATLAB / Simulink simulations, finite-element analysis, and experimental validation, we demonstrated that our improved method can steal over 76% of the power received by a fully resonant receiver under identical conditions. This attack demonstrates that simple frequency-changing power encryption offers limited protection against such threats.

## I. INTRODUCTION

WITH growing global concerns over the energy crisis and environmental pollution, the shift from fuel-powered vehicles to electric vehicles (EVs) has become an inevitable trend due to their high efficiency and zero emissions [1]. However, widespread adoption of EVs is hindered by limitations in battery technology, including high costs and low energy density [2]. To address these challenges, various solutions have been proposed, such as battery swap technology [3]. Yet, this automated battery replacement system faces significant limitations, as it is difficult to implement across different vehicle models.

Fortunately, advances in wireless power transfer (WPT) offer a promising solution: roadway wireless charging for EVs [4]. This innovation eliminates the hassle of long charging times and could eventually make battery-free EV possible [5, 6]. Therefore, it helps reduce the heavy metal pollution associated with battery production and improper recycling [7].

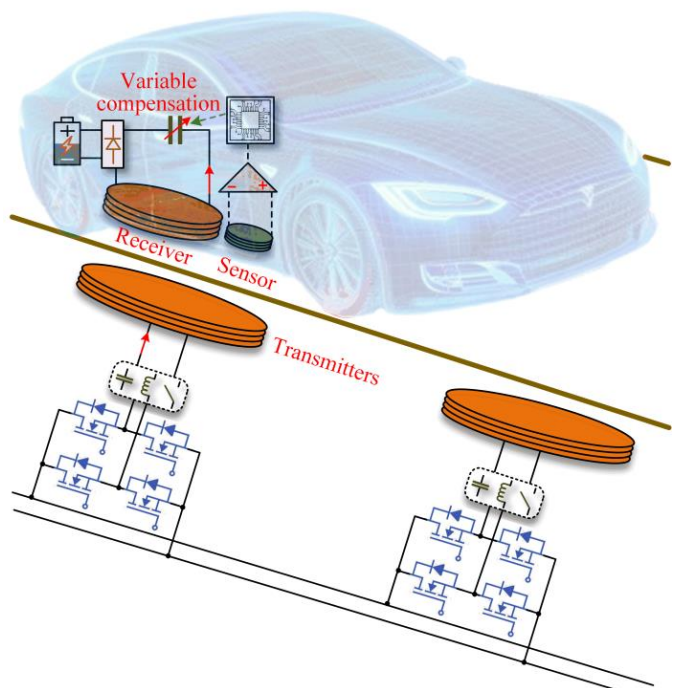

Fig. 1. WPT for roadway charging.

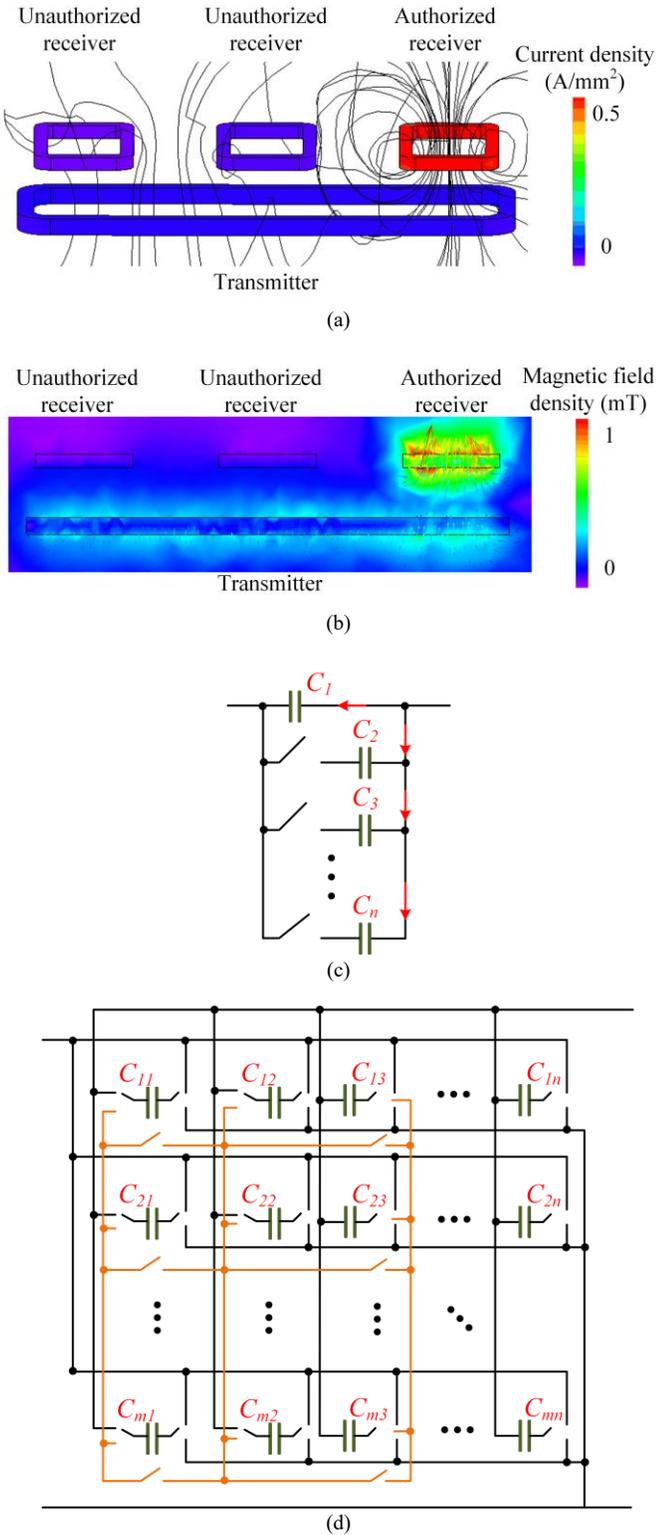

Fig. 2. Frequency-varying encryption strategy. (a) Magnetic field distribution of the energy encryption system. (b) Magnetic field density of the system. (c) Capacitor array. (d) Capacitor matrix.

However, this technology poses significant safety risks. Any device that has access to the magnetic field typically has the chance to extract wireless energy, whether it pays or not [8]. The energy thieves not only lead to financial losses for electricity providers but can even destabilize the power grid [4, 9-11].

Also, we have previously demonstrated that a hacker could calculate the frequency with a magnetic sensor coil and compensate the transmitter with a time-varying switched-capacitor circuit throughout a wide frequency range (Fig. 1) [12].

However, the previous exploit demonstration had a limited response speed or needed precise circuit parameters (inductances and capacitances) for detection and synchronization to a new frequency of about 100 milliseconds. Thus, the transmitter can use the fast frequency-varying strategy to protect energy safety [13, 14]. Only a resonant receiver with the right compensation can harvest energy efficiently, while non-resonant receivers suffer from a high impedance (Fig. 2(a) and (b)) [15]. Thus, the WPT frequency (sequence) serves as the key for this WPT link. To date, engineers have developed various compensation circuits for frequency-varying systems [16-19], such as capacitor arrays (Fig. 2(c)) [16] and topology-reconfigurable capacitor matrices (Fig. 2(d)) [20].

Therefore, readers could mistakenly believe that the fast frequency-varying encryption method is safe, especially with real-time low-latency modes as part of the development of the sixth generation (6G) of cellular communication technology, which is envisioned to precisely synchronize authorized users and rapidly distribute a new key [21].

In response, we redesigned our attack strategy and the associated setup from scratch to fill this gap. We demonstrate that our system can hack the wireless energy within even just three cycles, i.e., dozens of microseconds. Hence, there is no chance to use a fast frequency-hopping method to get rid of the energy hacker because the hacker knows the frequency as fast as or even faster than the communication synchronization between the transmitter and authorized receivers [22, 23].

Moreover, the new hacker can compensate for parameter drift in real time, whereas many authorized receivers may struggle with this issue, which is primarily caused by the aging of capacitors.

The key innovation is our use of phase detection technology instead of time-consuming receiver current measurements. This allows the hacker to calculate the initial duty cycle of the time-division switched capacitor within just three cycles of the WPT signal, and then refine it by synchronizing the receiver current with the phase of the magnetic sensor voltage. As a result, our new hacker demonstrates strong applicability in real-world scenarios, and the simple frequency-switching strategy (even with 5G and 6G communication) cannot ensure cybersecurity.

This paper is organized as follows: Section II presents the system configuration. Subsequently, Section III describes the system operation and designation procedures. Sections IV and V respectively demonstrate the system with simulations and experiments. Finally, Section VI summarizes the paper.

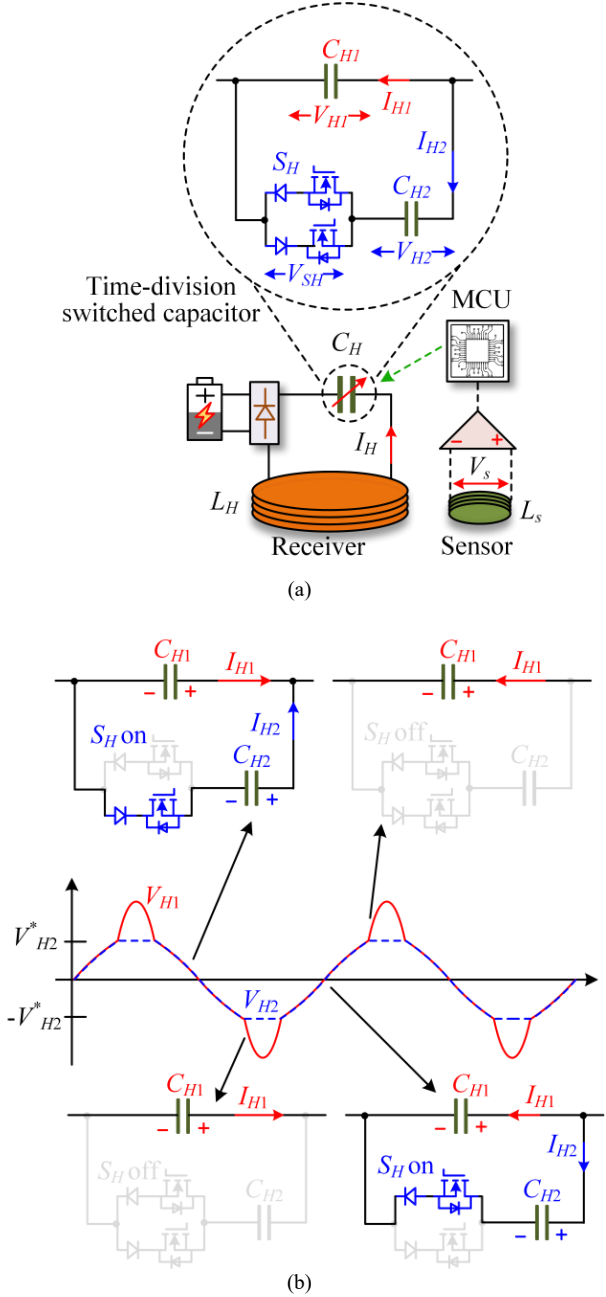

Fig. 3. The proposed hacking contraption. (a) Topology. (b) Working principle.

## II. Hacking System Topology and Operation

### A. System Topology

The structure of our hacker is simple. The larger power receiver coil $L_H$ can hack the power from magnetic fields (Fig. 3(a)); the variable capacitor $C_H$ can continuously compensate the hacking receiver $L_H$ in a wide frequency range; the small sensor coil $L_S$ works as a magnetic sensor to detect the frequency and phase information [24]; the rectifier bridge converts the AC power to DC power to charge the load battery. $V_{H1}$, $V_{H2}$, $V_{SH}$, $I_{H1}$, and $I_{H2}$ are the voltages and currents of the corresponding capacitors and the switch.

In contrast to previous frequency-varying systems [20], the variable capacitor $C_H$ only contains two capacitors ($C_{H1}$ and $C_{H2}$)

and one switch $S_H$. Also, different from other types of switches for controlled capacitors [25], $S_H$ contains two field-effect transistors (FETs) and two diodes. Even though the topology is more complicated, the control is simpler, as it can always achieve soft-switching control [12].

### B. Multi-Frequency Compensation Regulation

We assume that the magnetic fields of the transceiver use a wide frequency range [13]. So, the micro-controller unit (MCU) must keep regulating the capacitance $C_H$ to compensate $L_H$ and achieve resonance.

As in traditional LC circuits, the premise of resonance is that the capacitor and inductor can store the same energy. For the proposed time-division switched capacitor compensation, the relationship can be expressed as

$$P_{LH} = P_{CH1} + P_{CH2}$$
$$\Downarrow$$
$$\frac{1}{2} L_H I_H^2 = \frac{1}{2} C_{H1} V_{H1}^2 + \frac{1}{2} C_{H2} V_{H2}^2,$$

(1)

where $P_{LH}$, $P_{CH1}$, and $P_{CH2}$ are the power restored at $L_H$, $C_{H1}$, and $C_{H2}$, respectively.

Thus, the MCU can regulate $P_{CH2}$ by switching $S_H$ to control the $V_H^*$, i.e., maximum voltage of $C_{H2}$ in each cycle (Fig. 3(b)). Prominently, as the MCU can adjust the on-time of $S_H$ continuously, the hacker has the ability to resonate over a wide frequency range.

## III. Hacking Strategy

### A. Frequency and Phase Detection

Even though the frequency $f_T$ of the alternative magnetic field (AMF) is unknown and varying, we can detect it precisely and timely through a magnetic flux sensor $L_S$. Fast digital detection can time several upward zero crossings to easily measure the period and calculate the frequency $f_T$.

As Fig. 4(a) illustrates, we can get phase information of the magnetic field and calculate the frequency through the voltage of the sensor $V_S$, which in the first approximation follows [26]

$$V_S = j2\pi f_T M_{TS} I_T,$$

(2)

where $M_{TS}$ is the mutual inductance between $L_T$ and $L_S$.

Ideally, the overlap between $L_T$ and $L_H$ is small, so the stolen current does not affect the phase detection of $V_S$. Eq. (2) more accurately includes an interaction term per

$$V_S = j2\pi f_T M_{TS} I_T + j2\pi f_H M_{HS} I_H,$$

(3)

where $M_{HS}$ is the mutual inductance between $L_H$ and $L_S$; $f_H$ is the frequency of current $I_H$, which may be different from $f_T$ before compensation $C_H$ adjustment.

The phase of the stolen current $I_H$ plays a major role in regulating $P_{CH2}$. Even though one could in principle use high-frequency current sensors, they are costly and processing them with high bandwidth and low latency is complicated. Instead, we detect the load voltage $V_{HR}$ directly, as the load is purely resistive in this paper. Moreover, there is no need to use a complex digital-to-analog converter (ADC) module; only the zero-phase detection is enough for the MCU to learn the phase of $I_H$ (Fig. 4(b)).

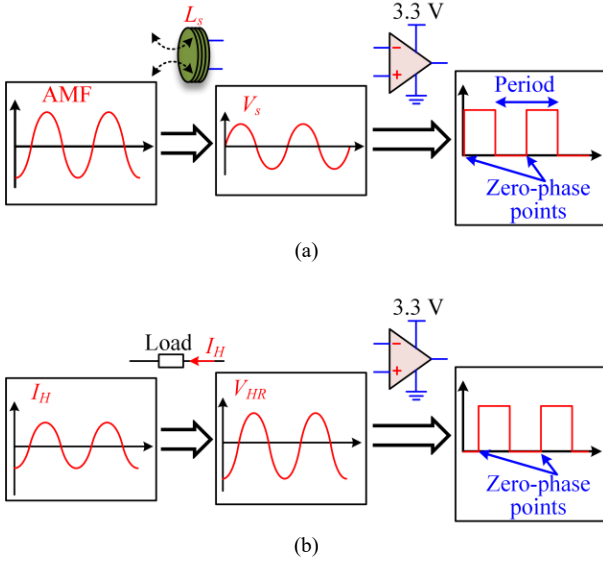

Fig. 4. Phase and frequency detection. (a) Magnetic field detection through sensor $L_S$. (b) Hacker current detection by detecting the load voltage.

### B. Compensation Regulation Strategy with All Parameters

As indicated above, the regulation process balances the energy stored in the inductive coil and capacitive compensation. For the hacking coil, the restored energy is easy to calculate as

$$P_{LH} = \frac{1}{2} L_H I_{Hmax}^2, \tag{4}$$

where $I_{Hmax}$ is the maximum current of the hacking circuit.

Instead of an ADC to detect $V_{H1}$ and $V_{H2}$ in real time and calculate the power through (1), which would massively load the controller, we instead use a timer to control $P_{CH1}$ and $P_{CH2}$ when we know the current waveform. In general, there are three conditions:

- Direct AC load (Fig. 5(a)). The hacking current $I_H$ is not sinusoidal (Fig. 5 (b)). The main reason is the tuning concept that temporally superpositions different impedances to achieve an average match over one period as well as resonance. Thus, the impedance varies during a period.
- DC load with rectifier (Fig. 5(c)). The current is not only non-sinusoidal but also discontinuous if $L_H$ is too small (Fig. 5(d)). The main reason is that the filter capacitor after the rectifier bridge has a certain DC voltage, which blocks current flow if the instantaneous input does not reach the forward voltage of the rectifier.
- DC load with rectifier and active power-factor correction (PFC) (Fig. 5(e)). We can control the current to be sinusoidal with a buck-type PFC circuit (Fig. 5 (f)).

For the third condition, the energy restored in $L_H$, $C_{H1}$, and $C_{H2}$ is easy to calculate according to (1), and the on-time $T_{ON}$ for each frequency point $f_H$ can be calculated through [12]

$$T_{ON} = \frac{1}{\pi f_T} \arcsin\left(\frac{C_{R1} + C_{R2}}{C_{R2}} - (2\pi f_T)^2 L_R \frac{C_{R1}(C_{R1} + C_{R2})}{C_{R2}}\right). \tag{5}$$

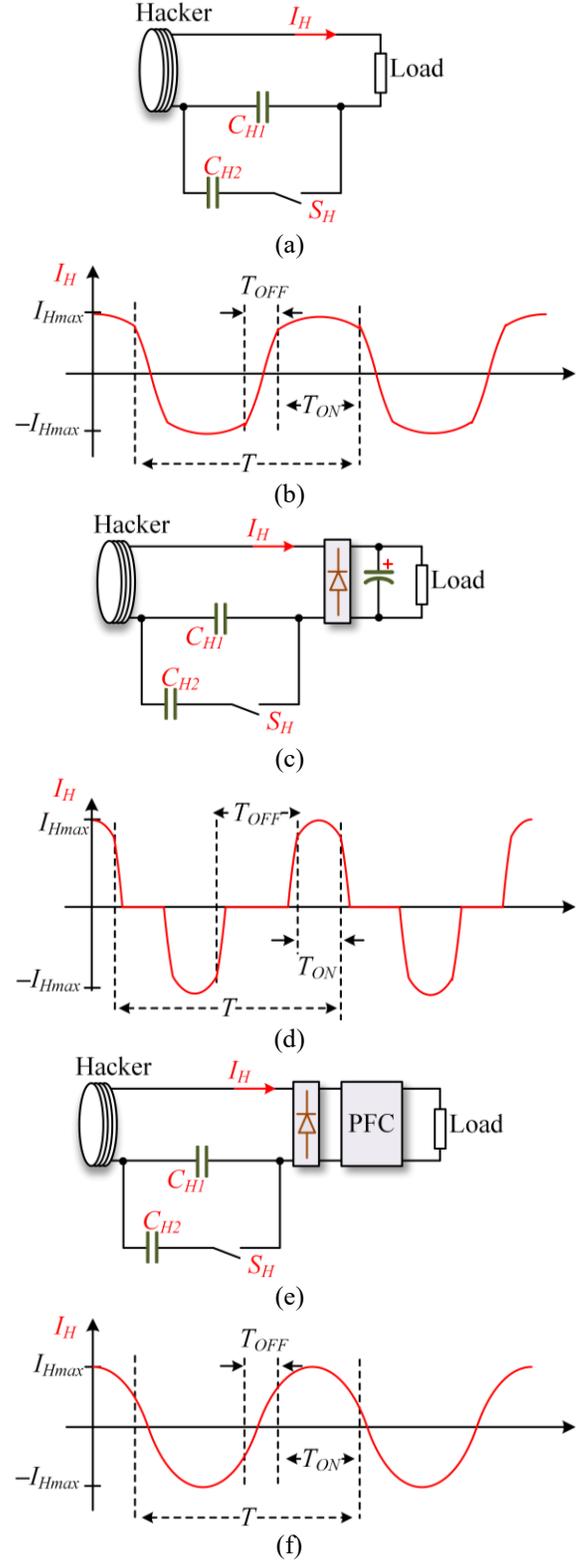

Fig. 5. Different hacking circuits and corresponding hacker currents. (a) Hacking circuit for direct AC load. (b) Continuous non-sinusoidal current. (c) Hacking circuit for the DC load with a rectifier. (d) Discontinuous current with a low-inductance receiver. (e) Hacking circuit for DC load with rectifier and PFC. (f) Continuous sinusoidal current.

However, for the first and second conditions, the current waveforms are piecewise functions. Especially in the second condition, the current is related to the voltage of the filter capacitor. Moreover, the precise values of $L_R$, $C_{R1}$, and $C_{R2}$ are hard to get and may drift over time. In these cases, this calculation method only offers an approximate value of $T_{ON}$ to initialize the tracking; after this initialization, we can use the phase control method to refine this value. The next subsection details this adaptive refinement or tracking of the switch timings.

### C. Compensation Regulation Strategy without Any Parameter

As we said before, this system can regulate compensation even without knowledge of fundamental parameters, such as $L_H$, $C_{H1}$, and $C_{H2}$. The solution tracks and systematically minimizes the phase difference between $V_S$ and $I_H$ or forces $V_S$ into leading $V_{H1}$ by 90°.

The main reason is that, when the hacking interceptor is resonant, $I_H$ would lead the transmitter current $I_T$ by 90°, while we already know $V_S$ leads the transmitter current $I_T$ by 90° according to (2). Thus, the relationship between these variables can be expressed as [27]

$$I_H = \frac{j2\pi f_T M_{TH}}{j2\pi f_T L_H + \frac{1}{j2\pi f_T C_H} + R_H} \times I_T$$

$$\Downarrow \left( \text{when } 2\pi f_T L_H = \frac{1}{2\pi f_T C_H} \right)$$

$$= \frac{j2\pi f_T M_{TH}}{R_H} \times I_T \qquad (6)$$

$$\Downarrow \quad V_S = j2\pi f_T M_{TS} I_T \ (2)$$

$$= \frac{M_{TH}}{R_H \times M_{TS}} V_S,$$

where $M_{TH}$ is the mutual inductance between transmitter coil $L_T$ and hacker $L_H$; $R_H$ denotes the equivalent load of the hacking interceptor circuit.

Therefore, as long as $I_H$ and $V_S$ share the same phase or $V_S$ leads $V_{H1}$ by 90°, the hacking circuit is resonant. In this paper, we use the voltage of the load $V_{HR}$ to get the phase of $I_H$. Fig. 4(b) details the phase tracking and tuning.

However, when $V_{HR}$ or $I_H$ leads $V_S$, it means the impedance of the hacking circuit is capacitive because (6) becomes

$$I_H = \frac{j2\pi f_T M_{TH}}{j2\pi f_T L_H + \frac{1}{j2\pi f_T C_H} + R_H} \times I_T$$

$$\Downarrow \left( \text{when } 2\pi f_T L_H < \frac{1}{2\pi f_T C_H} \right)$$

$$= \frac{j2\pi f_T M_{TH}}{R_H - jX_H} \times I_T \qquad (7)$$

$$\Downarrow \quad V_S = j2\pi f_T M_{TS} I_T \ (2)$$

$$= \frac{M_{TH}}{(R_H - jX_H) \times M_{TS}} V_S,$$

where $X_H = \frac{1}{2\pi f_T C_H} - 2\pi f_T L_H$, and it is a positive value in this case. Thus, the system only needs to increase $C_H$ to reduce $X_C$, and then the system will be resonant. Vice versa, when $V_{HR}$ or $I_H$ lags behind $V_S$, it means the impedance of the hacking circuit is inductive, and the system only needs to decrease $C_H$ by shortening $T_{ON}$. Therefore, the control logic can be simply represented by Fig. 6.

In contrast to our former attacks [12], there is no ADC in the energy hacking circuit. Because the hacker only detects the phase relationship of zero-phase points of $V_S$ and $V_{H1}$, which is substantially more efficient and faster than the older current-maximization regulation strategy.

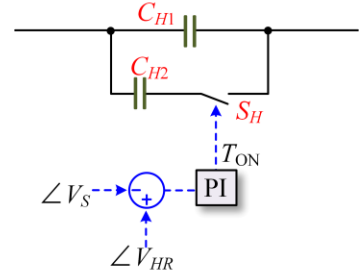

Fig. 6. Compensation regulation with a phase control strategy.

### D. The Overall Process of the Hacking System

Fig. 7 illustrates the hacking procedure in a flowchart. If the interceptor knows the key parameters, we can calculate the desired $T_{ON}$ through (8) first. Subsequently, we refine $T_{ON}$ through the given phase-control method.

If the hacking system is unaware of the key parameters, or the key parameters change for irrelevant reasons, e.g., aging, temperature, etc., we can directly let the phase-control method find and track $T_{ON}$.

Compared to former attempts, the new system significantly shortened the reaction time to three cycles. Also, the system even can hack energy without knowing the specific inductance of the hacking receiver coil or the capacitance of the compensating capacitors.

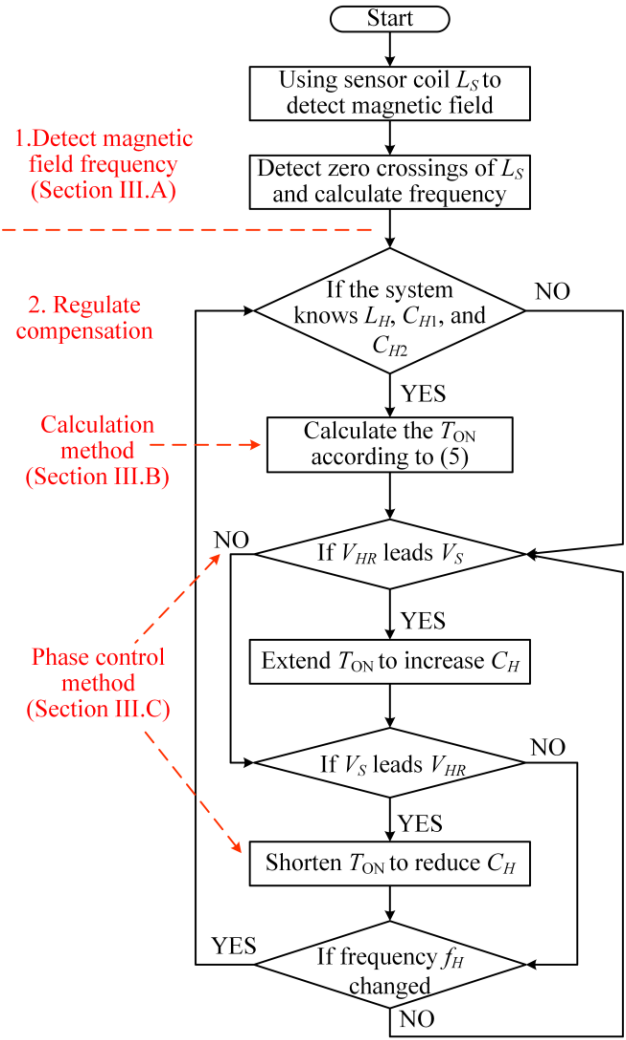

1. Detect magnetic field frequency (Section III.A)

2. Regulate compensation

Calculation method (Section III.B)

Phase control method (Section III.C)

Fig. 7. Flowchart of design and operation of the presented rapid interceptor.

## IV. SIMULATION

For verification, we analyzed the proposed new system with MATLAB / Simulink and JMAG.

### A. Circuit Simulation

In this MATLAB / Simulink simulation, the transmitter and hacking receiver coils $L_T$ and $L_H$ are respectively 140 µH and 93 µH; the compensation $C_{R1}$ and $C_{R2}$ are respectively 5 nF and 100 nF. The frequency range in which the receiver could track amounts to 51 kHz to 233 kHz.

Whereas previous demonstrations hacked energy with the knowledge of essential component parameters [12], we concentrate on stealing energy with the system's parameter-free mode.

When the transmitter generates 233 kHz, the system can be resonant within four cycles without the knowledge of any key parameters of the hacking system (Fig. 8(a)). Another three cycles later, the hacking current $I_H$ can be as high as the fully resonant receiver's current $I_R$. Thus, the hacking interceptor can work fast even solely through tracking the angle difference between the sensor voltage $V_S$ and the compensation voltage $V_{H1}$; the interceptor does not have to calculate $T_{ON}$ at all.

When the system operates at 51 kHz (Fig. 8(b)), the hacker needs a longer time to be fully resonant. Still, the system can steal as much power as the authorized receiver receives at no more than six cycles after the frequency change.

Thus, this simulation proves that the new hacking attack not only can steal energy over a wide frequency range but can also extract energy without the knowledge of key parameters.

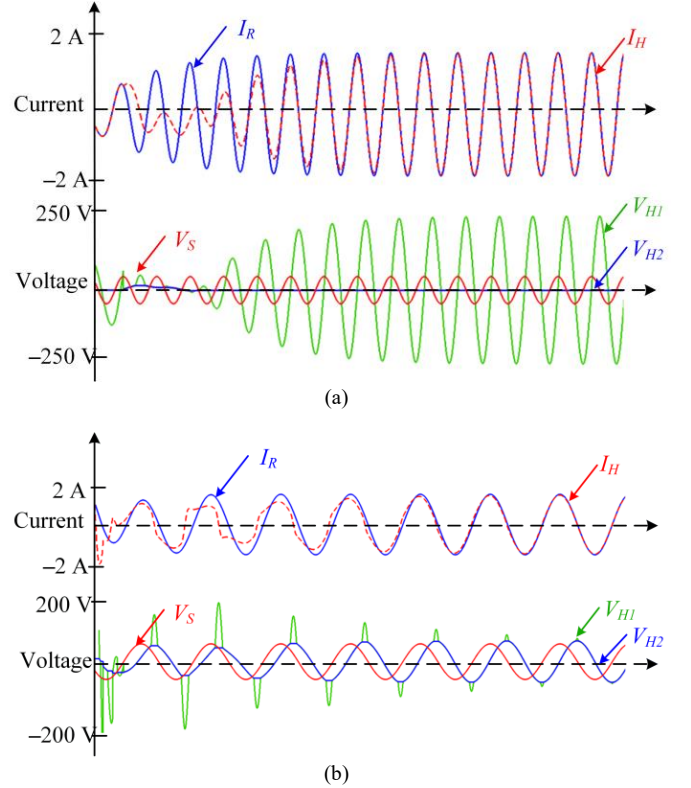

Fig. 8. Currents of the transceiver and the receivers at (a) 233 kHz and (b) 51 kHz.

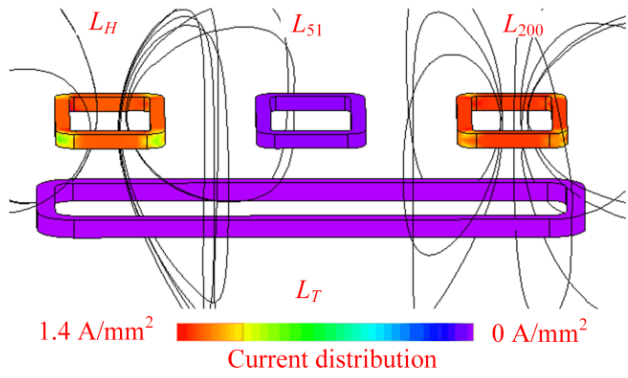

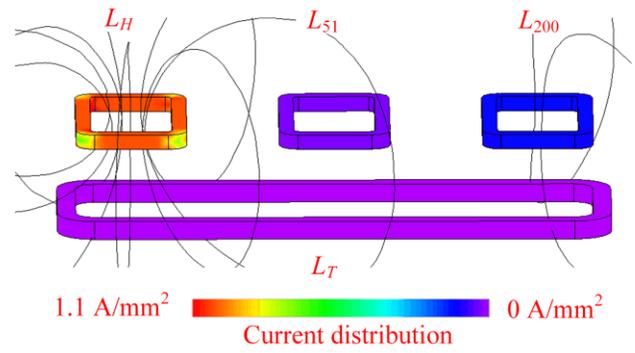

(a)

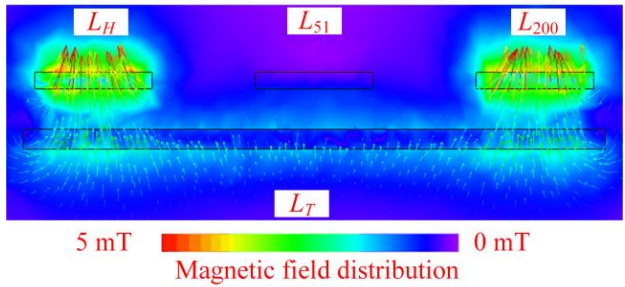

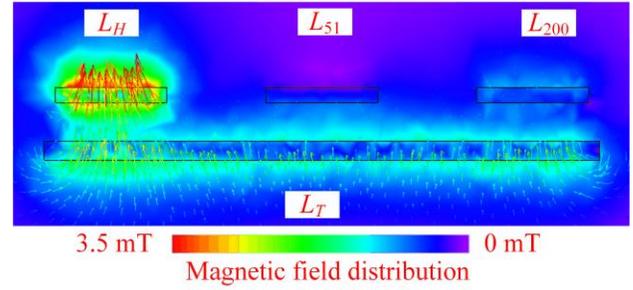

(b)

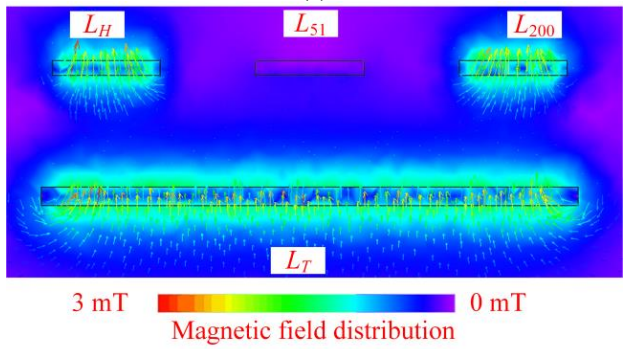

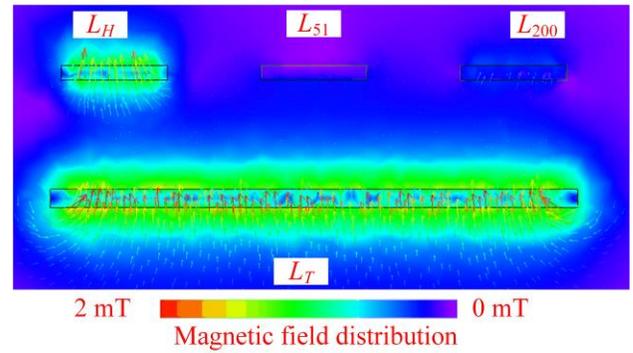

(c)

Fig. 9. Finite-element simulation results of the system at 200 kHz. (a) The current density and magnetic flux lines of each coil. (b) The system's magnetic flux distribution with a 20 mm transfer distance. (c) The system's magnetic flux distribution with a 60 mm transfer distance.

Fig. 10. Finite-element simulation results of the system at 120 kHz. (a) The current density and magnetic flux lines of each coil. (b) The system's magnetic flux distribution with a 20 mm transfer distance. (c) The system's magnetic flux distribution with a 60 mm transfer distance.

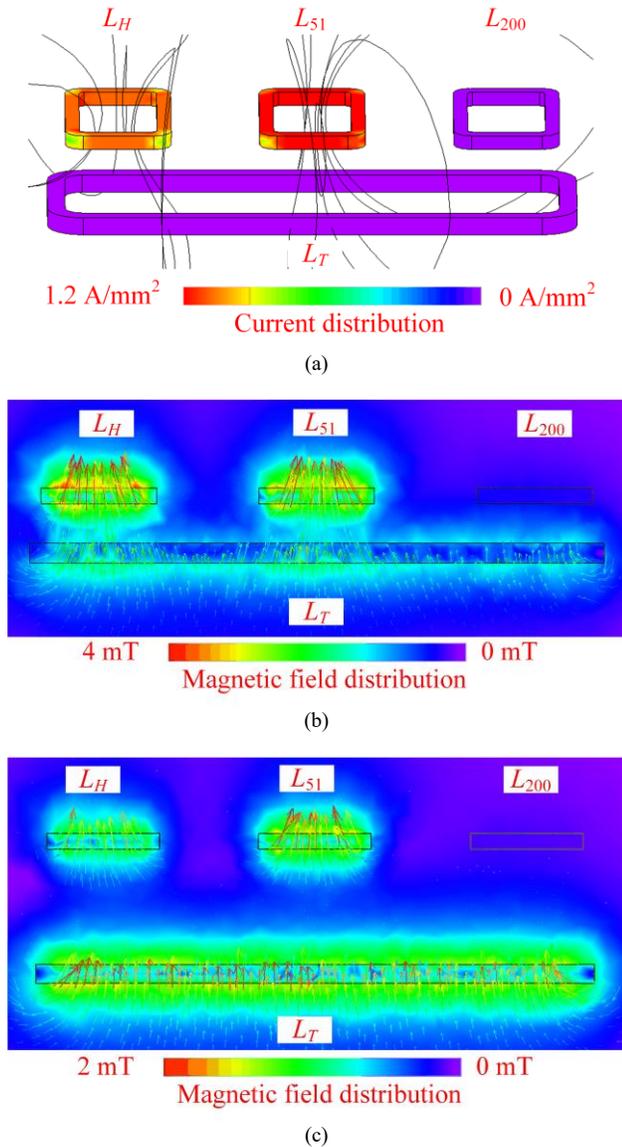

1.2 A/mm² [Current distribution] 0 A/mm²

(a)

4 mT [Magnetic field distribution] 0 mT

(b)

2 mT [Magnetic field distribution] 0 mT

(c)

Fig. 11. Finite-element simulation results of the system at 51 kHz. (a) The current density and magnetic flux lines of each coil. (b) The system's magnetic flux distribution with a 20 mm transfer distance. (c) The system's magnetic flux distribution with a 60 mm transfer distance.

## B. Finite-Element Simulation

We further used JMAG for finite-element simulation. For comparison, we added two more receivers $L_{51}$ and $L_{200}$, which are fully resonant respectively at 51 kHz and 200 kHz.

Even though three receivers are placed together, the magnetic flux and current densities of these three coils differ. Fig. 9 illustrates the power transfer at 200 kHz. Fixed-resonance $L_{51}$ can extract only very limited energy, whereas the hacker $L_H$ demonstrates almost the same magnetic flux and current densities as the fully resonant receiver $L_{200}$.

When $f_T$ hops to 120 kHz, only $L_H$ is resonant after regulation, while $L_{51}$ and $L_{200}$ are not. Thus, $L_H$ has a higher current density and higher magnetic field density accordingly (Fig. 10).

When $f_T$ changes to 200 kHz, $L_H$ can still resonate and extract almost as much energy as $L_{200}$, whereas $L_{51}$ receives practically nothing because of the high impedance (Figs. 11).

Also, further inspection of JMAG simulation results indicate that the hacking performance slightly decreases when the transfer distance increased from 20 to 60 mm, especially in Fig. 12(c). The main reason is that the hacking circuit has diodes and MOSFETs. When the transmission distance increases, the energy harvested by each receiving coil decreases, but the energy loss of the electronic device does not decrease significantly as the load power, so the hacking efficiency drops marginally.

TABLE I
KEY EXPERIMENT PARAMETERS

| Item | Value / Type | Unit |
|---|---|---|
| WPT frequency range ($f_T$) | 75–220 | kHz |
| Transmitter coil inductances ($L_T$) | 47 | μH |
| Inductances of hacker/receiver ($L_H$, $L_{R75}$, $L_{R220}$) | 37, 37, 37 | μH |
| Mutual inductance ($M_R$) | 6 | μH |
| Compensation capacitances ($C_{H1}$, $C_{H2}$) | 10, 122 | nF |
| Load resistances ($R_L$, $R_{75}$, $R_{220}$) | 10, 10, 10 | Ω |
| Wireless transfer distance | 20 | mm |
| Diameter of the transmitter coil (inner, outer) | 26, 28 | cm |
| Diameter of the receiver coil (inner, outer) | 13, 15 | cm |

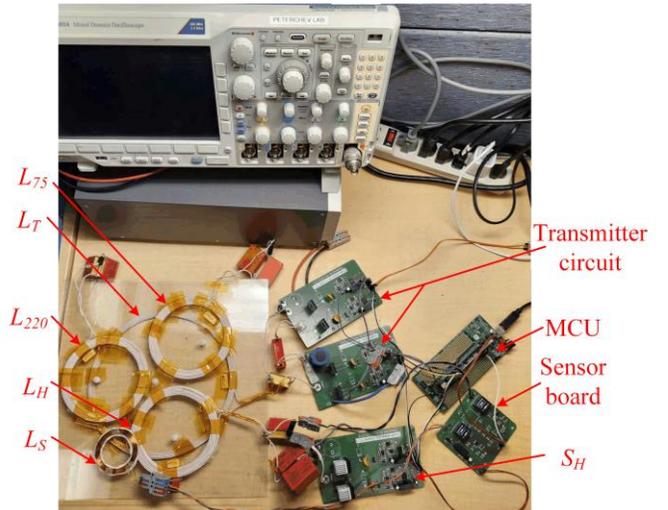

Fig. 12. Experimental setup.

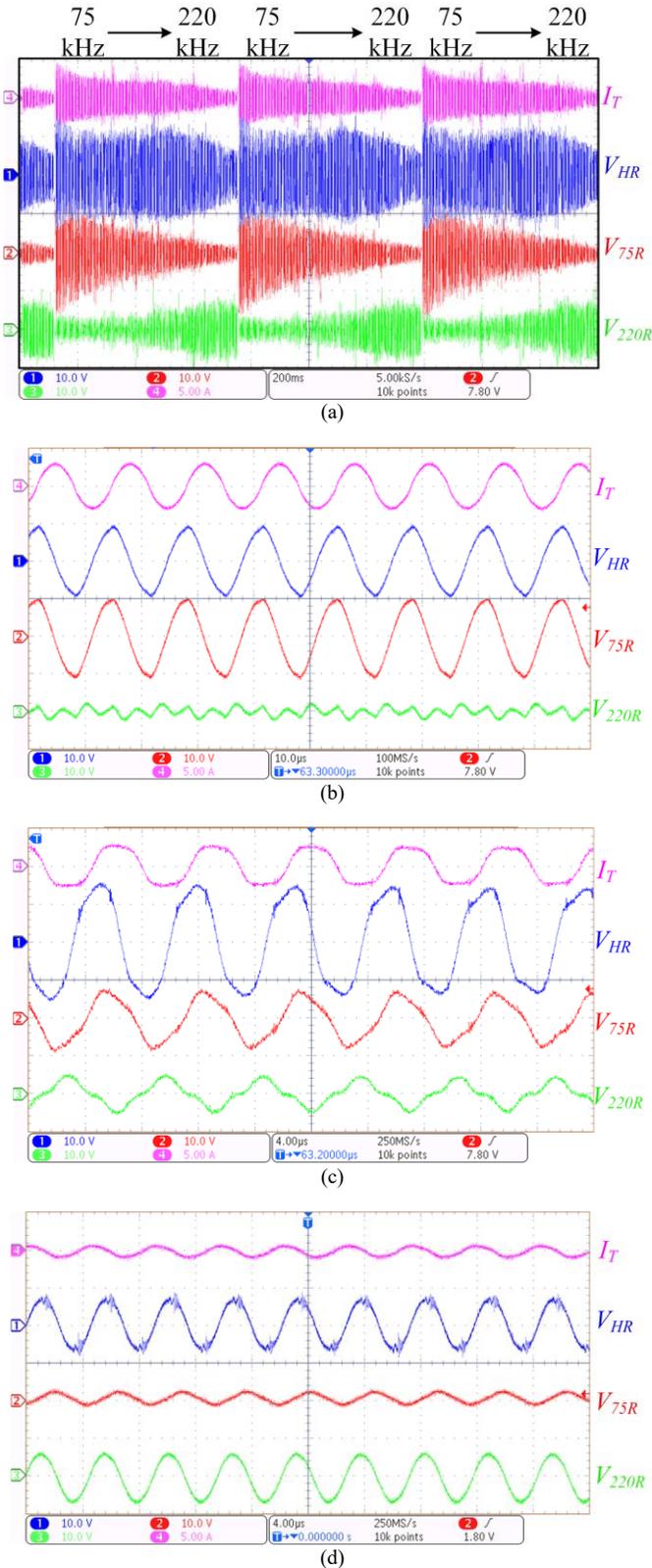

Fig. 13. Hacking performance at different frequency points. (a) From 75 to 220 kHz, (b) at 75 kHz, (c) at 150 kHz, (d) at 220 kHz.

## V. Experiments

We implemented an experimental setup to demonstrate the performance of the new energy decryption system (Fig. 12).

Similarly to the former system, there are a frequency-varying transmitter coil $L_T$, a power interceptor coil $L_H$, and the sensor coil $L_S$ (key parameters are given in Table I). We furthermore added two fixed-resonance receivers $L_{75}$ and $L_{220}$ for comparison, one at 75 kHz and one at 220 kHz.

When we sweep the transmitting frequency changes from 75 kHz to 220 kHz with 1 kHz increments, the voltage of the hacker's load $V_{HR}$ is always kept at least 5 V (Fig. 13). In comparison, the load voltage of the 75 kHz circuit, namely $V_{75R}$, decreases with the frequency, whereas the 220 kHz receiver can only effectively harvest energy at high frequency.

To be more specific, at 75 kHz, the rated voltages of $V_{HR}$, $V_{75R}$, and $V_{220R}$ are respectively 6.29 V, 7.19 V, and 0.3 V. Thus, the power ratio between these three loads is 0.76: 1: 0.002. At 150 kHz, $V_{HR}$, $V_{75R}$, and $V_{220R}$ change to 11.1 V, 5.32 V, and 3 V, respectively. Hence, the power ratio becomes 1: 0.23: 0.07. At 220 kHz, $V_{HR}$, $V_{75R}$, and $V_{220R}$ respectively become 4.6 V, 1.1 V, and 4.85 V. Therefore, the power ratio becomes 0.89: 0.05: 1. Therefore, the hacking efficiency can always be higher than 76% from 75 to 220 kHz. However, under the same condition, the nonresonant receiver can only extract 0.2% − 5% of the energy received by the properly resonant receiver. Certainly, as in the simulation results, the power ratio (hacking efficiency) is related to the power level, transfer distance, adopted components, and load condition. Those factors affect the ratio of the switch $S_H$ energy loss to load energy consumption. In other words, if the switch $S_H$ is well optimized, the hacking efficiency can be even higher than 76%.

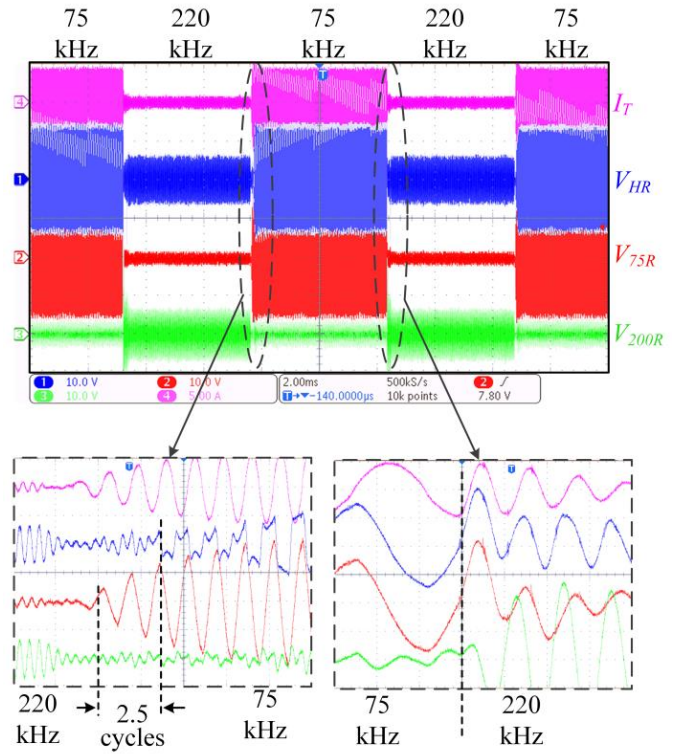

Fig. 14. The interceptor's ultra-fast tracking and transition speed when the frequency hops 155 kHz directly.

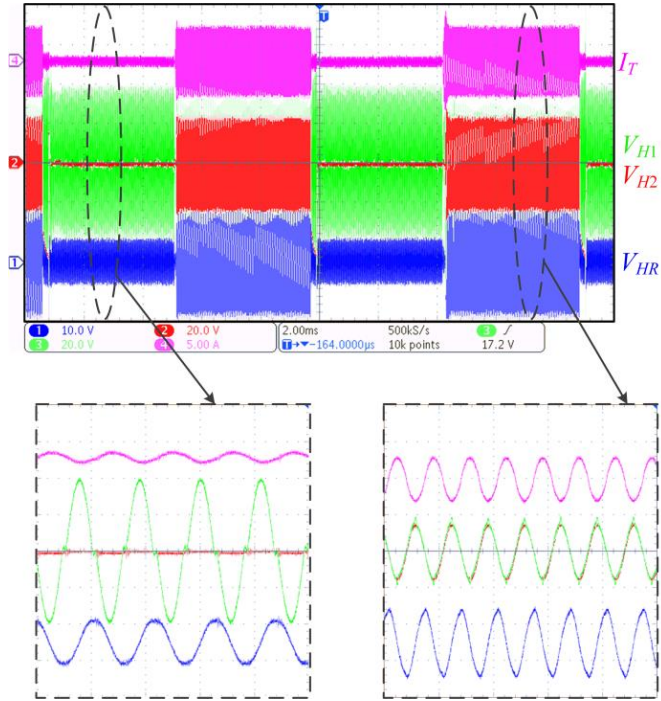

Fig. 15. The capacitors' voltages at different frequency points.

In contrast to the previous attack [12], our new interceptor can track the key and hack energy ultra-fast if the MCU knows all the circuit parameters. When the frequency change is very small, such as 1 kHz, we can hack it seamlessly without any latency or power interruption. Even if $f_T$ jumps from 220 kHz to 75 kHz, the interceptor can still respond fast. As shown in Fig. 14, the hacker is only about 2.5 cycles behind the resonant receiver. When $f_T$ jumps from 75 kHz to 220 kHz, the system can practically immediately harvest energy, even though the efficiency is not very high at the beginning.

Figure 15 charts the voltage waveforms of $C_{H1}$ and $C_{H2}$. The interceptor can control $P_{CH1}$ and $P_{CH2}$ effectively and efficiently.

## VI. Conclusion

This paper presented an ultra-fast WPT hacking system, which fills the gap in previous papers. Different from the former hacking demonstration, our new system abandons slow and not absolutely necessary ADC modules to regulate compensation directly through phase compensation. Therefore, the system can detect the frequency of the magnetic field and regulate the compensation within dozens of microseconds, while the former one needs at least 100 milliseconds. Prominently, the new hacker can overcome the parameter drift or even work without any system parameters. Moreover, the hacking efficiency, i.e., the stolen energy compared to an authorized receiver, can be up to 76%, which is basically the same as the former one (78%) and at least 15 times more efficient than nonresonant receivers in harvesting energy. These hacking efficiencies are achieved using a pure resistive load without a rectifier bridge or PFC. Thus, the results demonstrate that the proposed new hacking method achieves high efficiency and rapid response. Further active impedance control of the subsequent stage rather promised performance improvements.

Importantly, even small amounts of stolen energy can pose a significant threat to encrypted wireless power systems. Therefore, this new attack presents a considerable challenge to energy security and protection.

Whereas wireless power transfer and charging are rapidly winning over consumer electronics, professional and high-power applications—including static and particularly dynamic charging of electric cars for larger effective range—with publicly accessible charging points or tracks may strongly depend on means to protect it from unauthorized access to protect and justify the costly investment.

Power encryption could be an enabling technology, but it needs to be sufficiently strong. For example, the WPT system can transfer power and data together to concurrently authorize the user or to compare sent as well as received power. However, such communication may need to be bidirectional, require substantial hardware modifications, and cannot refer to any concept demonstration. Such a concept may furthermore only turn off the power transfer, but not protect it. An interceptor with access to the field, as we could show, can readily steal power anyway. Alternatively, the power supplier could try physical access protection, such as adding a dedicated lane for wireless charging and cameras for monitoring. These solutions may affect privacy and generate substantial capital costs for the additional dedicated access-controlled lanes. Due to sharp margins and increasing numbers of EVs, even small energy theft can break the economic case of protected wireless charging. In addition to the energy cost itself, it would have to be supplied by the infrastructure. Therefore, maintaining energy security is worth the investment.

In the future, our work will focus on hacking dual-frequency or multi-frequency WPT systems. Ideally, an interceptor would be able to identify the strongest magnetic field in multiple frequencies to steal that energy. A potential technical approach may use a series of phase-locked loops to simultaneously detect multiple frequencies of the current magnetic field. Artificial intelligence could select the most suitable power source and regulate the compensation to steal energy.

## Author Contributions

HW performed the research, implemented the method, and wrote the manuscript. HDS secured funding and advised the research team. SMG designed the research, secured funding, and wrote the manuscript.